\documentclass[12pt,twoside]{article}
\usepackage{correl,mathrsfs} 
\usepackage[T1]{fontenc}

\def\TheTitle{Geometry and topology \\ in many-body physics}
\def\TheHeading{Geometrical observables}
\def\TheAuthor{Raffaele Resta}
\def\TheInstitute{Istituto Officina dei Materiali IOM-CNR, Trieste, Italy}
\def\TheAddress{and Donostia International Physics Center, San Sebasti{\'a}n, Spain}
\def\TheChapter{}                       

\newcommand{\ei}[1]{{\rm e}^{i #1}}
\newcommand{\emi}[1]{{\rm e}^{-i #1}}
\newcommand{\kk}{\mbox{\boldmath$\kappa$}}

\newcommand{\da}{\partial_{k_\alpha}}
\newcommand{\db}{\partial_{k_\beta}}

\def\beq{\begin{equation}}
\def\eeq{\end{equation}}
\def\bea{\begin{eqnarray}}
\def\eea{\end{eqnarray}}
\def\nn{\nonumber\\}

\def\ket#1{\vert#1\rangle}

\def\me#1#2#3{\langle#1\vert#2\vert#3\rangle}
\def\ev#1{\langle#1\rangle}
\newcommand{\daa}{\partial_{\kappa_\alpha}}
\newcommand{\dbb}{\partial_{\kappa_\beta}}

\def\EE{{\cal E}}
\newcommand{\GG}{{\cal G}}
\def\EEE{\mbox{\boldmath${\cal E}$}}
\def\r{{\bf r}}
\def\v{{\bf v}}
\def\k{{\bf k}}

\def\p{{\bf p}}
\def\R{{\bf R}}

\def\M{{\bf M}}

\def\FF{{\cal F}}
\def\GG{{\cal G}}
\def\P{{\bf P}}

\def\P0{ \Psi_0(0)}

\renewcommand{\[}{\begin{equation}}
\renewcommand{\]}{\end{equation}}
\renewcommand{\Large}{\normalsize}

\newcommand{\equ}[1]{Eq.~(\ref{#1})}
\newcommand{\eqs}[2]{Eqs.~(\ref{#1}) and (\ref{#2})}

\def\runtime{(\the\time)\qquad\the\month/\the\day/\the\year}
\def\today
 {\count10=\year\advance\count10 by -2000 \number\day--\ifcase
  \month \or Jan\or Feb\or Mar\or Apr\or May\or Jun\or
             Jul\or Aug\or Sep\or Oct\or Nov\or Dec\fi--\number\count10}

\def\hour{\count10=\time\count11=\count10
\divide\count10 by 60 \count12=\count10
\multiply\count12 by 60 \advance\count11 by -\count12\count12=0
\number\count10 :\ifnum\count11 < 10 \number\count12\fi\number\count11}
\def\draft1#1{
\setlength{\unitlength}{1cm}
\noindent \begin{picture}(0,0)
\put(0,2.5){\noindent \sf DRAFT: {\bf #1} run through \LaTeX\ on \today\ at \hour}
\end{picture}
}
\begin{document}

\MakeTitle           
\section{Introduction}

Some intensive observables of the electronic ground state in condensed matter have a geometrical or even topological nature. In crystalline systems at the noninteracting (or mean-field) level the term ``geometrical'' refers to the geometry of the occupied manifold of the state vectors, parametrized by the Bloch vector $\k$ in reciprocal space. A state-of-the-art account about several of such observables can be found in the recent outstanding book by D. Vanderbilt \cite{Vanderbilt}. 

In the present Review I present, instead, the known geometrical observables beyond band-structure theory, in order to deal with the general case of disordered and/or correlated many-electron systems. The term ``geometrical'' refers therefore to an Hilbert space (defined below in Sect. \ref{sec:hilbert}) different from the Bloch space. 

It is now clear that the geometrical observables come in two very different classes.
The observables of class (i) only make sense for insulators, and are defined modulo $2\pi$ (in dimensionless units), while the observables of class (ii) are defined for both insulators and metals, and are single-valued. 

As for class (i), two observables are known: electrical polarization and the ``axion'' term in magnetoelectric response \cite{Vanderbilt}.  For both observables the modulo $2\pi$ ambiguity is fixed only after the termination of the insulating sample is specified. Furthermore in presence of some protecting symmetry only the values zero or $\pi$ (mod $2\pi$) are allowed: the observable becomes then a topological  ${\mathbb Z}_2$ index. So far, the expression of the axion term is only known within band-structure theory: therefore in the present Review I only discuss electrical polarization, whose many-body expression was first obtained in 1998 \cite{rap100}. The geometrical nature of polarization is thoroughly investigated in Sect. \ref{sec:pol}, while in Sect. \ref{sec:topo} it is shown that 1$d$ polarization in inversion-symmetric systems is a
${\mathbb Z}_2$ invariant. 

After discussing polarization, I will address four observables of class (ii); they do not include the case of orbital magnetization, whose geometrical expression is known since 2006 at the band-structure level \cite{rap130}, but which to date lacks a corresponding many-body formulation. Of these four observables two are time-reversal (T) even and two are T-odd; the latter are nonzero only if the material breaks T-symmetry. The T-even are the Drude weight and the Souza-Wilkens-Martin sum rule; the T-odd ones are the anomalous Hall conductivity and the magnetic circular dichroism sum rule. It may appear surprising that I include spectral sum rules in the class of ground-state observables: this is because, owing to a fluctuation-dissipation theorem, a frequency-integrated dynamical probe becomes effectively a static one. The corresponding physical property cannot be actually measured with a static probe, but is nonetheless a genuine ground-state property. All of the four single-valued observables---despite being ground-state properties---have to do with the conductivity tensor $\sigma_{\alpha\beta}(\omega)$; therefore, before addressing them, in Sect. \ref{sec:conductivity} I display the full many-body Kubo formul\ae. They comprise four terms: real and imaginary, symmetric (longitudinal) and antisymmetric (transverse). 

The content of Sects. \ref{sec:even}  and \ref{sec:odd} is a thorough discussion of the four class-(ii) geometrical observables and of their consequences, in particular for the theory of the insulating state. A synoptic view of all five observables object of this Review is provided in the concluding Sect. \ref{sec:conclu}. Some boring derivations are confined to the Appendix.

\section{What does it mean ``geometrical'' in quantum mechanics?} \label{sec:geometry}

The funding concept in geometry is distance. Let $\ket{\Psi_1}$ and $\ket{\Psi_2}$ be two quantum states in the same Hilbert space: it is expedient to adopt for their pseudodistance the expression \[ {\cal D}^2_{12} = - \ln |\ev{\Psi_1 | \Psi_2}|^2 .\label{b1} \] It is ``pseudo'' because it violates one of the distance axioms in calculus textbooks; such violation does not make any harm in the present context.

\equ{b1} vanishes when the states $\ket{\Psi_1}$ and $\ket{\Psi_2}$ coincide, while it diverges when the states $\ket{\Psi_1}$ and $\ket{\Psi_2}$ are orthogonal. The states $\ket{\Psi_1}$ and $\ket{\Psi_2}$ are defined up to an arbitrary phase factor: fixing this factor amounts to a gauge choice. \equ{b1} is clearly gauge-invariant. 

The distance in \equ{b1} can equivalently be rewritten as \[ {\cal D}^2_{12} = - \ln \ev{\Psi_1 | \Psi_2}  - \ln \ev{\Psi_2 | \Psi_1} , \label{b2} \] where the two terms are not separately gauge-invariant. While the distance is obviously real, each of the two terms in \equ{b2} is in general a complex number. If we write \[ \ev{\Psi_1 | \Psi_2} = |\ev{\Psi_1 | \Psi_2}| \ei{\varphi_{21}}, \] then the imaginary part of each of the two terms in \equ{b2} assumes a transparent meaning: \[ - \mbox{Im ln } \ev{\Psi_1 | \Psi_2} = \varphi_{12} , \qquad \varphi_{21} = - \varphi_{12}. \label{b3} \]
Besides the metric, an additional geometrical concept is therefore needed: the connection, which fixes the relative phases betweeen two states in the Hilbert space.

The connection is arbitrary and cannot have any physical meaning by itself. Nonetheless, after the 1984 groundbreaking paper by Michael Berry \cite{Berry84}, several physical observables are expressed in terms of the connection and related quantities.
\index{connection} \index{metric} \index{curvature}
When the state vector is a differentiable function of some parameter $\kk$, then the differential phase and the differential distance define the Berry connection and the quantum metric, respectively: 
\bea \varphi_{\kk,\kk+d\kk} = {\cal A}_\alpha(\kk) d \kappa_\alpha, & D_{\kk,\kk+d\kk}^2 = g_{\alpha\beta}(\kk)d\kappa_\alpha d\kappa_\beta , \label{conne} \\ {\cal A}_\alpha(\kk) = i \ev{\Psi_{\kk} | \partial_{\kappa_\alpha} \Psi_{\kk}}, \quad &
g_{\alpha\beta}(\kk) = \mbox{ Re } \ev{\partial_{\kappa_\alpha} \Psi_{\kk} | \partial_{\kappa_\beta} \Psi_{\kk}} - \ev{\partial_{\kappa_\alpha} \Psi_{\kk} | \Psi_{\kk} } \ev{\Psi_{\kk} | \partial_{\kappa_\beta} \Psi_{\kk}} ; \label{metric} \eea summation over repeated Cartesian indices is understood (here and throughout). The Berry curvature is defined as the curl of the connection: 
\[{\sf \Omega}_{\alpha\beta}(\kk)d\kappa_\alpha d\kappa_\beta  = [ \partial_ {\kappa_\alpha} {\cal A}_\beta(\kk) - \partial_{\kappa_\beta} {\cal A}_\alpha(\kk) ]d\kappa_\alpha d\kappa_\beta =  -2\,\mbox{Im} \ev{\da \Psi_{\kk} | \db \Psi_{\kk}}d\kappa_\alpha d\kappa_\beta  . \label{curva} \] The connection is a 1-form and is gauge-dependent; the metric and the curvature are 2-forms and are gauge invariant. The above fundamental quantities are defined in terms of the state vectors solely; we will also address a 2-form which involves the Hamiltonian as well. Suppose that $H$ is the Hamiltonian and $E_0$ its ground eigenvalue: we will consider \[ {\cal G} = \me{\Psi}{ ( H- E_0 ) }{\Psi}, \label{G} \]  which vanishes for $\ket{\Psi} = \ket{\Psi_0}$; an essential feature of ${\cal G}$ is that it is invariant by translation of the energy zero. The geometrical quantity of interest is the gauge-invariant 2-form which obtains by varying $\ket{\Psi}$ in the neighborhood of $\ket{\Psi_0}$.

\section{Many-body geometry} \label{sec:hilbert}

We address here the geometry of the many-body state vectors by generalizing the Hilbert space defined by W. Kohn in a milestone paper published in 1964 \cite{Kohn64}, well before any geometrical or topological concepts entered condensed matter physics. 

For the sake of simplicity we deal with the simple case where a purely orbital Hamiltonian can be established. Following Kohn, we consider a system of $N$ interacting $d$-dimensional electrons in a cubic box of volume $L^d$, and the family of many-body Hamiltonians parametrized by the parameter $\kk$: \[ \hat{H}_{\kk} = \frac{1}{2m} \sum_{i-1}^N \left[\p_i + \frac{e}{c} {\bf A}(\r_i) + \hbar \kk \right]^2 + \hat{V}, \label{kohn} \] where $\hat{V}$ includes one-body and two-body potentials. We assume the system to be macroscopically homogeneous; the eigenstates $\ket{\Psi_{n\kk}}$ are normalized to one in the hypercube of volume $L^{Nd}$. The vector potential ${\bf A}(\r)$ summarizes all T-breaking terms, as e.g. those due to spin-orbit coupling to a background of local moments. The vector $\kk$, having the dimensions of an inverse length, is called ``flux'' or ``twist'' and amounts to a gauge transformation. In order to simplify notations we will set $\hat{H}_{0} \equiv \hat{H}$, $\ket{\Psi_{n0}} \equiv \ket{\Psi_{n}}$ , $E_{n0} \equiv E_n$. \index{Kohn}

Bulk properties of condensed matter obtain from the thermodynamic limit: $N \rightarrow \infty$, $L \rightarrow \infty$, $N/L^d$ constant. All of the observables discussed here include $\kk$-derivatives of the state vectors $\ket{\Psi_{n\kk}}$: it is important to stress that the differentiation is performed first, and the thermodynamic limit afterwards. This ensures that a given eigenstate is followed adiabatically while the flux is turned on.
Kohn's Hamiltonian can be adopted within two different boundary conditions, thus defining two different Hilbert spaces.

\subsection{Open-boundary-conditions Hilbert space}

Within the so-called ``open'' boundary conditions (OBCs) one assumes that the cubic box confines the electrons in an infinite potential well; we will indicate as $\ket{\tilde{\Psi}_{n\kk}}$ the OBCs eigenstates, square-integrable over ${\mathbb R}^{Nd}$. Within OBCs the effect of the gauge is easily ``gauged away'': the energy eigenvalues $E_n$ are gauge-independent, while the eigenstates are $\ket{\tilde{\Psi}_{n\kk}} = \emi{\kk \cdot \hat\r} \ket{\tilde{\Psi}_n}$, where  $\hat\r = \sum_i \r_i$ is the many-body position (multiplicative) operator, well defined in this Hilbert space. 

\subsection{Periodic-boundary-conditions Hilbert space}

Within Born-von-K\`arm\`an periodic boundary conditions (PBCs) one assumes that the many-body wavefunctions are periodic with period $L$ over each electron coordinate $\r_i$ independently, whose Cartesian components
$r_{i,\alpha}$ are then equivalent to the angles $2\pi r_{i,\alpha}/L$. The
potential $\hat{V}$ and the vector potential ${\bf A}(\r)$ enjoy the same periodicity: this means that the macroscopic $\EEE$ and ${\bf B}$ fields vanish. It is worth observing that the position $\hat{\r}$ is {\it not} a legitimate operator in this Hilbert space: it maps a vector of the space into something which does not belong to the space \cite{rap100}.

As said above, setting $\kk \neq 0$ amounts to a gauge transformation; since PBCs violate gauge-invariance, the eigenvectors $\ket{\Psi_{n\kk}}$ and the eigenvalues $E_{n\kk}$ have a nontrivial $\kk$-dependence \cite{Kohn64}. 
The macroscopic ground-state current density  is \[ {\bf j}_{\kk} = - \frac{e}{\hbar L^d} \me{\Psi_{0\kk}}{\partial_{\kk} \hat{H}_{\kk}}{\Psi_{0\kk}} = - \frac{e}{\hbar L^d} \partial_{\kk} E_{0\kk} \; ; \label{current} \] it vanishes at any $\kk$ in insulators;\footnote{A mobility gap implies that any infinitesimal perturbation to the Hamiltonian does not induce a macroscopic current.} within OBCs it vanishes even in metals

An important comment is in order. Here we follow Kohn, by keeping the boundary conditions fixed and ``twisting'' the Hamiltonian; other authors \cite{Niu85} have addressed the many-body geometry by keeping the Hamiltonian fixed, and ``twisting'' the boundary conditions. The equivalence between the two approaches is rather straightforward.

\section{Macroscopic electrical polarization} \label{sec:pol}

Macroscopic electrical polarization only makes sense for insulators which are charge-neutral in average, and is comprised of an electronic (quantum) term and a nuclear (classical) term. Each of the terms separately depends on the choice of the coordinate origin, while their sum is translationally invariant; we also assume that the system is T-invariant, such that all $\kk = 0$ wavefunctions are real. \index{polarization}

\subsection{Bounded samples within open boundary conditions}

We consider, for the time being, the electronic term only. Within OBCs the observable has a pretty trivial definition: \[ {\bf P}^{(\rm el)} = -\frac{e}{L^d} \me{\tilde\Psi_0}{\hat{\r}}{\tilde\Psi_0} . \label{molec} \] I am going to transform \equ{molec} into a geometric form:
using $\ket{\tilde{\Psi}_{0\kk}} = \emi{\kk \cdot \hat\r} \ket{\tilde{\Psi}_0}$, one gets  \[ {\bf P}^{(\rm el)} = \frac{i e}{L^d} \ev{\tilde\Psi_0| \partial_{\kk} \tilde\Psi_0} = - \frac{e}{L^d} \mbox{\boldmath${\cal \tilde{A}}$}(0) . \label{molec2} \] The Berry connection is gauge dependent and cannot express a physical observable per se; we have in fact arrived at \equ{molec2} by enforcing a specific gauge. The most general $\kk$-dependence of the state vector is
$\ket{\tilde{\Psi}_{0\kk}} = \emi{\vartheta(\kk,\hat{\r})} \ket{\tilde{\Psi}_0}$, where $\vartheta(\kk,\hat{\r})= \kk\cdot \hat{\r} +\phi(\kk)$ where the gauge function $\phi(\kk)$ is arbitrary; \equ{molec2} makes sense only if we impose a gauge which makes $\vartheta(\kk,\hat{\r})$ odd in $\hat{\bf\r}$ at any $\kk$. 

\subsection{Unbounded samples within periodic boundary conditions} \index{connection}

We may try to adopt within PBCs the same definition as in \equ{molec2}: \[ {\bf P}^{(\rm el)} = \frac{i e}{L^d} \ev{\Psi_0| \partial_{\kk} \Psi_0} = - \frac{e}{L^d} \mbox{\boldmath${\cal A}$}(0), \label{sol} \] an obviously gauge-dependent expression. If, for instance, we evaluate the $\kk$-derivative by means of perturbation theory: \[ \ket{\partial_{\kk} \Psi_0} =  \sum_{n \neq 0}\ket{\Psi_n} \frac{\me{\Psi_n}{\partial_{\kk} \hat{H}}{\Psi_0}}{E_0 - E_n}, \label{pert}\] then we get $\ev{\Psi_0| \partial_{\kk} \Psi_0}=0$. In fact the parallel-transport gauge is implicit in the standard perturbation formula.
\index{boundary conditions}
In order to fix the gauge in a similar way as we did in the OBCs case, we realize that
$\emi{\kk \cdot \hat\r} \ket{{\Psi}_0}$ in general does not belong to the Hilbert space, bar in the cases where the $\kk$ components are integer multiples of $2\pi/L$. It is easy to verify that in such cases $\emi{\kk \cdot \hat\r} \ket{{\Psi}_0}$ is the ground eigenstate of $\hat{H}_{\kk}$ with eigenvalue $E_0$. We choose a $\kk$ in this set:
\[ \kk_1 = (2\pi/L, 0, 0).\] Since the connection is by definition the differential phase, \eqs{b3}{conne} yield to leading order \[ {\cal A}_x(0) \frac{2\pi}{L} \simeq - \mbox{Im ln } \ev{\Psi_0|\Psi_{0\kk_1}} , \qquad {\cal A}_x(0)  \simeq - \frac{L}{2\pi}\mbox{Im ln } \ev{\Psi_0|\Psi_{0\kk_1}} ; \label{scale} \]
\equ{sol} yields \[  P_x^{(\rm el)} = \frac{e}{2\pi L^{d-1}} \mbox{Im ln } \ev{\Psi_0|\Psi_{0\kk_1}} . \label{rap1} \] The state $\ket{\Psi_{0\kk}}$ is by definition the eigenstate of $\hat{H}_{\kk}$ which obtains by following $\ket{\Psi_{0}}$ adiabatically while the flux $\kk$ is turned on; owing to \equ{current}, its energy in insulators is $E_0$ ($\kk$-independent). Therefore in insulators---and in insulators {\it only}---$\ket{\Psi_{0\kk1}}$ is the ground eigenstate of $\hat{H}_{\kk_1}$; we fix its gauge by choosing $\ket{\Psi_{0\kk_1}}=\emi{\kk_1 \cdot \hat\r} \ket{{\Psi}_0}$, in the same way as we did in the OBCs case:
\[  P_x^{(\rm el)} = \frac{e}{2\pi L^{d-1}} \mbox{Im ln } \me{\Psi_0}{\emi{\kk_1 \cdot \hat\r}}{\Psi_0} = - \frac{e}{2\pi L^{d-1}} \mbox{Im ln } \me{\Psi_0}{\ei{\frac{2\pi}{L} \sum_i x_i}}{\Psi_0} . \label{rap2} \]  The polarization is intensive, ergo the logarithm scales like $N^{1-1/d}$, while the modulus of its argument tends to one from below. It is worth observing that the present gauge choice can be regarded as the many-body analogue of the periodic gauge in band-structure theory \cite{Vanderbilt}: see \equ{gauge} below, and the related footnote.
\equ{rap2} is the so-called single-point Berry-phase formula \cite{rap100}; for a crystalline system of noninteracting electrons it yields the (by now famous) Berry-phase formula in band-structure theory \cite{rap155}, first obtained by King-Smith and Vanderbilt in 1993 \cite{Vanderbilt,King93} (see also the Appendix).

When the Hamiltonian is adiabatically varied $\ket{\Psi_0}$ acquires an adiabatic time-dependence. It can be proved that $j^{(\rm el)}_x$, defined as  \[ j^{(el)}_x = \dot{P}^{(\rm el)}_x = \frac{e}{2\pi L^{d-1}} \mbox{Im } \left( \frac{\me{\dot\Psi_0}{\emi{\kk_1 \cdot \hat\r}}{\Psi_0}}{\me{\Psi_0}{\emi{\kk_1 \cdot \hat\r}}{\Psi_0}} + \frac{\me{\Psi_0}{\emi{\kk_1 \cdot \hat\r}}{\dot\Psi_0}}{\me{\Psi_0}{\emi{\kk_1 \cdot \hat\r}}{\Psi_0}} \right) , \] coincides indeed---to leading order in $1/L$---with the adiabatic current density which traverses the sample \cite{rap100,rap155}.

The nuclear term can be elegantly included in \equ{rap2}. If $X_\ell$ is the $x$ coordinate of the $\ell$-th nucleus with charge $eZ_\ell$, then \[  P_x = - \frac{e}{2\pi L^{d-1}} \mbox{Im ln } \me{\Psi_0}{\ei{\frac{2\pi}{L}(\, \sum_i x_i - \sum_\ell Z_\ell X_\ell\,)}}{\Psi_0} , \label{rap3} \] clearly invariant by translation of the coordinate origin. This expression also applies if the quantum nature of the nuclei is considered, and $\ket{\Psi_0}$ includes the nuclear degrees of freedom.

\subsection{Multivalued nature of polarization} \index{Berry phase}

We define the single-point Berry phase $\gamma_x$, including the nuclear contribution, as \[ \gamma_x = \mbox{Im ln } \me{\Psi_0}{\ei{\frac{2\pi}{L}(\, \sum_i x_i - \sum_\ell Z_\ell X_\ell\,)}}{\Psi_0}, \qquad P_x = - \frac{e}{2\pi L^{d-1}} \gamma_x . \label{rap4} \] After \equ{scale}, the single-point Berry phase scales like $N^{1-1/d}$.
Given that  $\gamma_x$ is arbitrary modulo $2\pi$, bulk polarization within PBCs is a multivalued vector. This may appear a disturbing mathematical artefact, but is instead a key feature of the real world. In the following we analyze separately three different cases: 1$d$ systems, 3$d$ crystalline systems, and 3$d$ noncrystalline systems at the independent-electron level.

\subsubsection{One-dimensional polarization}

The polarization $P$ of a quasi-1$d$ system (e.g. a stereoregular polymer) has the dimensions of a pure charge; in the unbounded case within PBCs $P$ is arbitrary modulo $e$. The modulo ambiguity is fixed only after the sample termination is specified: we are going to show this in detail on the paradigmatic example of polyacetylene, where the Berry phase yields $P=0$ mod $e$.

\begin{figure}[t] 
\begin{minipage}[b]{.65\linewidth}
\includegraphics[width=9.5cm]{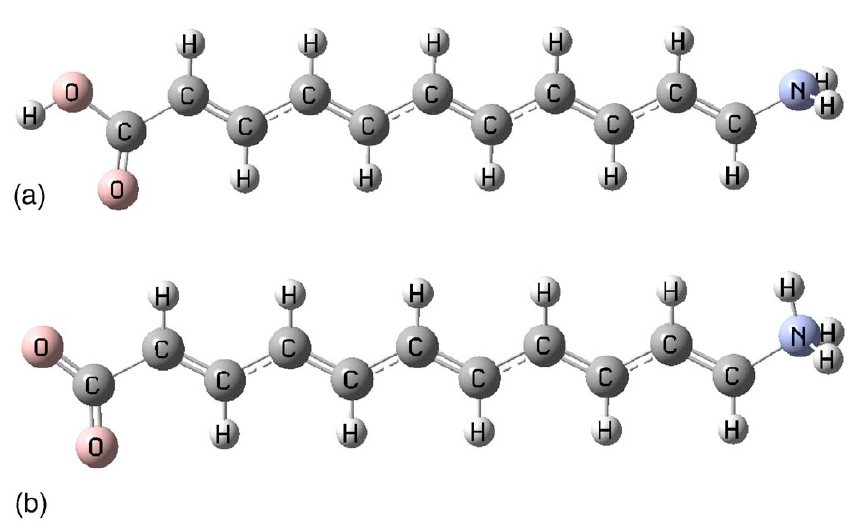}
\end{minipage}\hfill 
\begin{minipage}[b]{.35\linewidth} 
\caption{\rm A centrosymmetric polymer with two different terminations: alternant trans-polyacetylene. Here the ``bulk'' is five-monomer long. After Ref. \cite{rap136}. \vspace{1.5cm} ~~
\label{fig:quantum1}} 
\end{minipage}
\end{figure} 

\begin{figure}[t] 
\begin{minipage}[b]{.60\linewidth}
\includegraphics[width=8cm]{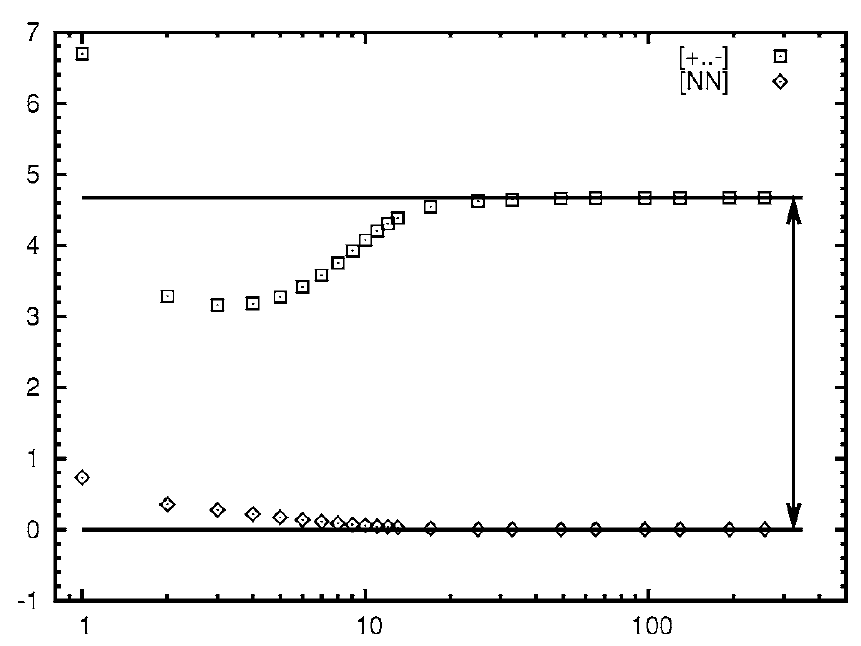}
\end{minipage}\hfill 
\begin{minipage}[b]{.40\linewidth}
\caption{\rm Quantization of polarization in polyacetylene: dipole per monomer (a.u.) as a function of the number of monomers in the chain, for the two different terminations. After Ref. \cite{rap136}. \vspace{2cm} ~~
\label{fig:quantum2}} 
\end{minipage}
\end{figure} 

We consider two differently terminated samples of trans-polyacetylene, as shown in Fig. \ref{fig:quantum1}: notice that in both cases the molecule as a whole is {\it not} inversion symmetric, although the bulk is. The dipoles of such molecules have been computed for several lengths from the Hartree-Fock ground state, as provided by a standard quantum-chemistry code \cite{rap136}. The dipoles per monomer are plotted in in Fig. \ref{fig:quantum2}: for small lengths both dipoles are nonzero, as expected, while in the large-chain limit they clearly converge to a quantized value. Since the lattice constant is $a=4.67$ bohr, the dipole per unit length is $P=0$ and $P=e$ for the two cases. The results in Fig. \ref{fig:quantum2} are in perspicuous agreement with the Berry-phase theory: in the two bounded realizations of the same quasi one-dimensional periodic system the dipole per unit length assumes---in the large-system limit---two of the values provided by the theory. Insofar as  the system is unbounded the modulo $e$ ambiguity in the $P$ value cannot be removed. 

\subsubsection{Three-dimensional crystalline polarization}

In the 3$d$ case \equ{rap4} yields \[ P_x = - \frac{e}{2\pi L^2} \gamma_x  \label{rap5} , \] which clearly cannot be used as it stands in the $L \rightarrow \infty$ limit. Notwithstanding, polarization is a well defined multivalued observable whenever the system is {\it crystalline}: with this we mean that a uniquely defined lattice can be associated with the real sample. The lattice is an abstraction, which is uniquely defined even in cases with correlation, quantum nuclei, chemical disorder---i.e. crystalline alloys, a.k.a. solid solutions---where the actual wavefunction may require a supercell (multiple of the primitive lattice cell).

For the sake of simplicity we consider---without loss of generality---a simple cubic lattice of constant $a$. The supercell side $L$ is an integer multiple of $a$: $L=Ma$. The integral is over a 3$N$-dimensional hypercube of sides $L$: \[ \me{\Psi_0}{\ei{\frac{2\pi}{L}(\, \sum_i x_i - \sum_\ell Z_\ell X_\ell\,)}}{\Psi_0}  = \int_{\rm hcube} \prod_{i=1}^N d \r_i  \; \ei{\frac{2\pi}{L}(\, \sum_i x_i - \sum_\ell Z_\ell X_\ell\,)} |\ev{\r_1,\r_2\dots\r_N|\Psi_0}|^2 . \] Owing to the crystalline hypothesis, the integral is equal the sum of $M^2$ identical integrals: see the Appendix for a proof. Therefore we may define a reduced matrix element and a reduced Berry phase \[ \tilde\gamma_x = \mbox{Im ln } \frac{1}{M^2} \int_{\rm hcube} \prod_{i=1}^N d \r_i  \; \ei{\frac{2\pi}{L}(\, \sum_i x_i - \sum_\ell Z_\ell X_\ell\,)} |\ev{\r_1,\r_2\dots\r_N|\Psi_0}|^2 , \label{reduced} \] in terms of which \[ P_x = - \frac{e}{2\pi a^2} \tilde{\gamma}_x . \] the polarization of a crystal is therefore a well defined multivalued crystalline observable, ambiguous modulo $e/a^2$ in each Cartesian component in the case of a simple cubic lattice. 

A generic lattice is dealt with by means of a coordinate transformation \cite{rap_a12}; the bulk value of ${\bf P}$ is then ambiguous modulo $e\R/V_{\rm cell}$, where $\R$ is a lattice vector and $V_{\rm cell}$ is the volume of a primitive cell. By definition a primitive cell is a minimum-volume one: this choice is mandatory in order to make ${\bf P}$ a well defined multivalued observable. As in the 1$d$ case, the modulo ambiguity is resolved only after the sample termination is specified; there are same complications, though. The theory, owing to PBCs and to the hypothesis of macroscopic homogeneity, yields the polarization ${\bf P}$ in zero $\EEE$ field; instead shape-dependent depolarization fields are generally present in a polarized 3$d$ sample. The depolarization field is zero for a sample in the form of a slab, and with ${\bf P}$ parallel to the slab (transverse case) \cite{rap_a30}. The second complication is the possible occurrence of metallic surfaces. Both complications are ruled out in the quasi-1$d$ case discussed above.

\subsubsection{Infrared spectra of liquid and amorphous systems}

Whenever a lattice cannot be defined, \equ{rap5} shows that ${\bf P}$ itself is not a ground-state observable in the thermodynamic limit. Nonetheless the single-point Berry phase of \equ{rap5}, at finite size $L$, is instrumental for evaluating polarization {\it differences}, or macroscopic currents; the latter are the key entry in the theory of infrared spectra. It is enough to choose $L$ larger than the relevant correlation lengths in the material; \equ{rap5} can then be used to access polarization differences $\Delta {\bf P}$ much smaller than $e/L^2$. \index{infrared spectroscopy}

For the sake of completeness we show here the form of \equ{rap5} when $\ket{\Psi_0}$ is the Slater determinant of $N/2$ doubly occupied $\k=0$ (supercell-periodical) Kohn-Sham orbitals $\ket{u_j} = \ket{\psi_j}$. One defines the connection matrix \[ S_{jj'} = \me{u_j}{\ei{\frac{2\pi}{L}x}}{u_{j'}} ; \] by including the nuclei and accounting for double orbital occupancy the polarization, in terms of the instantaneous Kohn-Sham orbitals, is \[ P_x(t) = - \frac{e}{2\pi L^2} \gamma_x = - \frac{e}{2\pi L^2} \mbox{Im ln } \left[ \, (\mbox{det } S)^2 \emi{\frac{2\pi}{L}\sum_\ell Z_\ell X_\ell} \, \right] .\]

The key quantity in the infrared spectra is the imaginary part of the isotropic dielectric response. The Kubo-Greenwood formula yields \[ \varepsilon"(\omega) = \frac{2\pi \omega}{3 L^3 k_{\rm B} T} \int_{-\infty}^\infty dt \; \langle \, {\bf d}(t) \cdot {\bf d}(0) \, \rangle , \] where ${\bf d} = L^3 {\bf P}$ is the dipole of the simulation cell and the brackets indicate the thermal average. In a Car-Parrinello simulation the integrand is evaluated at discrete time steps, and only small polarization differences are needed: at any discretized time $n\Delta t$ the polarization is \bea {\bf P}(n\Delta t)  &=& {\bf P}(0) + [ {\bf P}(\Delta t) - {\bf P}(0) ] + [ {\bf P}(2\Delta t) - {\bf P}(\Delta t) ] + \dots \nn &+& [ {\bf P}(n\Delta t) - {\bf P}((n-1)\Delta t). \eea  Not surprising, the material whose infrared spectrum has been most studied is liquid water. The very first  Car-Parrinello infrared spectrum for liquid water appeared in 1997 \cite{Silvestrelli97}; many other followed over the years.

\section{Topological polarization in one dimension} \label{sec:topo}

In presence of inversion symmetry $P=-P$, ergo either $P=0$ or $P=e/2$, mod $e$. This feature has clearly a one-to-one mapping to ${\mathbb Z}_2$, the additive group of the integers modulo two. The polarization of a centrosymmetric polymer is in fact topological: one cannot continuously transform a  ${\mathbb Z}_2$-even insulator into a ${\mathbb Z}_2$-odd---by enforcing inversion symmetry---without passing through a metallic state.
Arguably, this is the simplest occurrence of a ${\mathbb Z}_2$ topological invariant in condensed matter physics. Similar arguments lead to the quantization of the soliton charge in polyacetylene, whose topological nature was discovered by Su, Schrieffer, and Heeger back in 1979 \cite{Su79}; they also considered more generally non-singlet cases (here we always assume a nondegenerate singlet ground state). 

Fig. \ref{fig:quantum2} shows that quantization occurs in the large-$L$ limit only: this is an OBCs feature. Within PBCs quantization occurs even at finite $L$: in all inversion symmetric cases, the matrix element in \equ{rap3} is always real: either positive (${\mathbb Z}_2$-even) or negative (${\mathbb Z}_2$-odd).

The above results clearly demonstrate that polyacetylene is a ${\mathbb Z}_2$-even topological case. A paradigmatic ${\mathbb Z}_2$-odd case instead is a one-dimensional ``ionic crystal'': a linear chain of alternating equidistant anions and cations. In the long-chain limit $P=e/2$ mod $e$, independently of the ionicity of the two atoms; this happens e.g. for the two-band Hubbard model discussed next, at low $U$ values.

A topological quantum transition---occurring in a paradigmatic highly correlated system---was identified long ago in Refs. \cite{rap87} and \cite{rap107}, although no topological jargon was in fashion at the time. Here I reinterpret topology-wise the original results.

The model system addressed was the two-band Hubbard model (at half filling):
\[ H \! = \! \sum_{j \sigma} [  (-1)^j \Delta \,
c^\dagger_{j \sigma} c_{j \sigma} - t ( c^\dagger_{j \sigma} c_{j+1 \sigma} +
\mbox{H.c.} ) ] + \, U \sum_j n_{j\uparrow}  n_{j\downarrow}  . \label{hub} \] We assume $\Delta>0$, and neutralizing classical charges equal to $+1$ on all sites; the system is clearly inversion-symmetric at any $U$.

\begin{figure}[b] 
\begin{minipage}[b]{.60\linewidth}
\includegraphics[width=9cm]{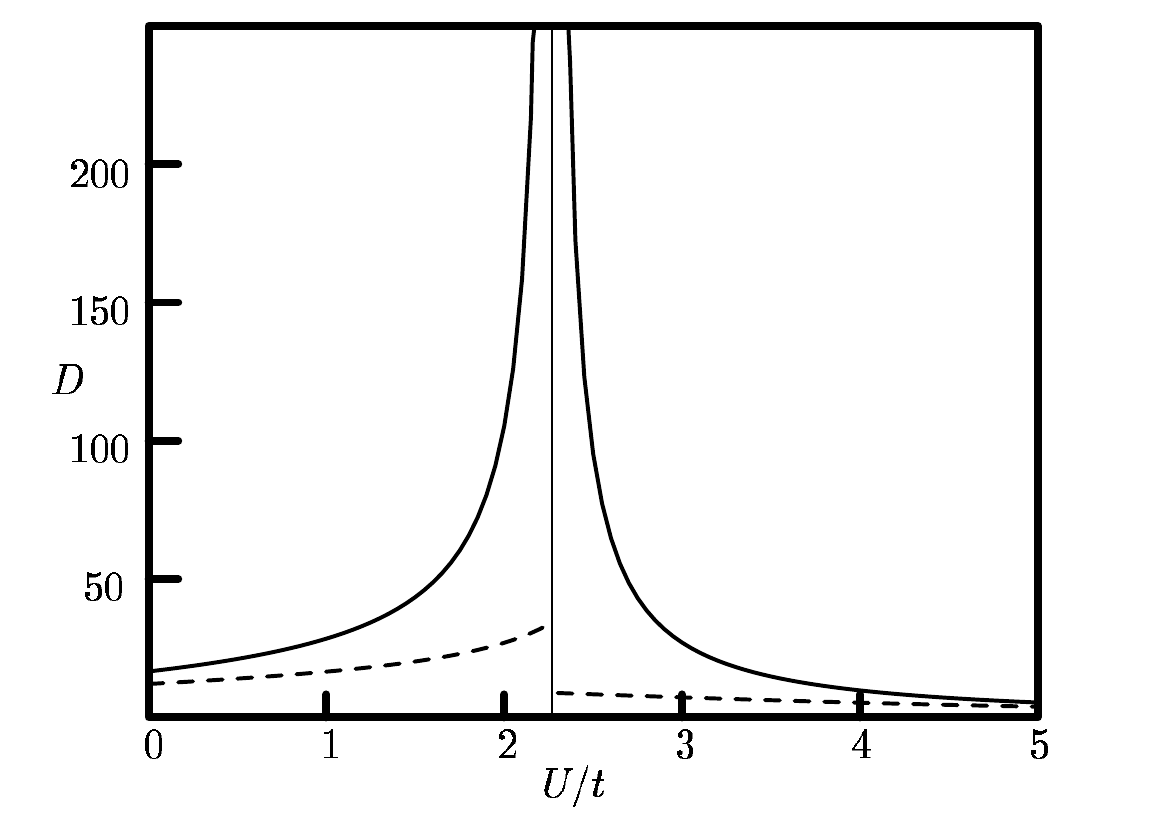}
\end{minipage}\hfill 
\begin{minipage}[b]{.40\linewidth}
\caption{\rm Squared localization length for the Hamiltonian in \protect\equ{hub} at half filling for $t/\Delta = 1.75$: the plot shows the dimensionless quantity $D=(2\pi N/L)^2 \lambda^2$. The system undergoes a quantum phase transition from band-like insulator (${\mathbb Z}_2$-odd) to Mott-like insulator (${\mathbb Z}_2$-even) at $U/t=2.27$. After Ref. \protect\cite{rap107}.\vspace{1cm}
\label{fig:hubbard}}
\end{minipage}
\end{figure} 

Preliminarly, it is expedient to investigate the trivial $t=0$ case. At small $U$ the anion site (odd $j$) is doubly occupied, and the energy per cell is $-2\Delta +U$; at $U > 2\Delta$ single occupancy of each site is instead energetically favored. As for polarization, it is easily realized that the system is ${\mathbb Z}_2$-odd in the former case and ${\mathbb Z}_2$-even in the latter. At the transition point $U_c = 2\Delta$ the ground state is degenerate and the spectrum is gapless, ergo the system is ``metallic''. If the hopping $t$ is then switched on adiabatically, the ${\mathbb Z}_2$ invariant in each of the two topological phases cannot flip unless a metallic state is crossed.

\index{topological invariant} \index{Hubbard}
Finite $t$ simulations have been performed in Ref. \cite{rap107} for several $U$ values,
where the explicitly correlated ground-state wavefunction has been found by exact diagonalization, at fixed $t/\Delta = 1.75$. The insulating/metallic character of the system was monitored by means of the squared localization length  \[ \lambda^2 =   -\frac{L^2}{4\pi^2 N} \mbox{ln } |\me{\Psi_0}{\ei{\frac{2\pi}{L} \sum_i x_i}}{\Psi_0}|^2  \label{rs} , \]  which will be addressed in detail in Sect. \ref{sec:rs} below. For the time being, suffices to say that in the large-$N$ limit $\lambda^2$ stays finite in all kinds of insulators while it diverges in metals.

The results of the simulations are shown in Fig. \ref{fig:hubbard}. The $t=0$ arguments presented above guarantee that at low $U$ values the system is a band-like insulator (${\mathbb Z}_2$-odd) , while at high $U$ values it is a Mott-like insulator (${\mathbb Z}_2$-even). The sharp transition occurs at the singular point $U_c=2.27t$;
there is no metal-insulator transition, only an insulator-insulator transition, while the system is metallic at the transition point. If we start  from the pure band insulator at $U=0$, there is a single occupied band and a doubly occupied Wannier function, centered at the anion site: therefore $P=e/2$ mod $e$ \cite{Vanderbilt}. Suppose now we switch on the Hubbard $U$ continuously: the Wannier function is no longer defined, while polarization $P$, \equ{rap3},  is well defined at any $U$ value ($U_c$ excepted). At the transition point the gap closes and $P$ flips to the 0 (mod $e$) value for $U>U_c$.

Remarkably, it was found that the static ionic charges (on anion and cation) are continuous across the transition, while they are instead obviously discontinuous in the $t=0$ case. It was also found that the the dynamical (Born) effective charge on a given site changes sign \cite{rap87} at the transition; in retrospect, we now understand that such sign change in a linear-response property was indeed the fingerprint of the flip of the topological ${\mathbb Z}_2$ index in the ground state.

\section{Kubo formul\ae\ for conductivity} \label{sec:conductivity}

Although this review only concerns ground-state properties, it is expedient to display the whole Kubo formul\ae\ for the dynamical conductivity $\sigma_{\alpha\beta}(\omega)$.
We define the $\kk=0$ many-body velocity operator and its matrix elements: \[ \hat{\v} = \frac{1}{\hbar} \partial_{\kk} \hat{H} = \frac{1}{m} \sum_{i-1}^N \left[\p_i + \frac{e}{c} {\bf A}(\r_i) \right]  
\]
 \[ {\cal R}_{n,\alpha\beta} = \mbox{Re }\langle \Psi_0 | \hat v_\alpha | \Psi_n \rangle \langle
\Psi_n | \hat v_\beta | \Psi_0 \rangle  , \quad  {\cal I}_{n,\alpha\beta} = \mbox{Im }\langle \Psi_0 | \hat v_\alpha | \Psi_n \rangle \langle \Psi_n | \hat v_\beta | \Psi_0 \rangle ,
\label{vmat} \] where ${\cal R}_{n,\alpha\beta}$ is symmetric and ${\cal I}_{n,\alpha\beta} $ antisymmetric; we further set $\omega_{0n} = (E_n - E_0)/\hbar$.
The longitudinal (symmetric) conductivity is: \[  \sigma_{\alpha\beta}^{(+)}(\omega) = D_{\alpha\beta} \left[ \delta(\omega) + \frac{i}{\pi \omega} \right] +\sigma_{\alpha\beta}^{(\rm regular)}(\omega) , \label{cond} \] 
\index{conductivity}
\[ D_{\alpha\beta} = \frac{\pi e^2}{ L^d} \left( \frac{N}{m} \delta_{\alpha\beta} - \frac{2}{\hbar} {\sum_{n\neq 0}}  \frac{{\cal R}_{n,\alpha\beta}  }{\omega_{0n}} \right) \label{drude} , \]
\bea \mbox{Re } \sigma_{\alpha\beta}^{(\rm regular)}(\omega) &=& \frac{\pi e^2}{\hbar L^d}  {\sum_{n\neq 0}} \frac{ {\cal R}_{n,\alpha\beta}}{\omega_{0n}} [\, \delta(\omega - \omega_{0n}) + \delta(\omega + \omega_{0n}) \,] , \label{s1} \\ \mbox{Im } \sigma_{\alpha\beta}^{(\rm regular)}(\omega) &=& \frac{2 e^2}{\hbar L^d}  {\sum_{n\neq 0}} \frac{ {\cal R}_{n,\alpha\beta}}{\omega_{0n}} \frac{\omega}{\omega_{0n}^2 - \omega^2} \label{s2} . \eea It will be shown below that the Drude weight $D_{\alpha\beta}$ can be regarded as a geometrical property of the many-electron ground state; it vanishes in insulators. The real part of longitudinal conductivity obeys the $f$-sum rule  \[ \int_0^\infty d \omega \; \mbox{Re } \sigma_{\alpha\beta} (\omega) = \frac{D_{\alpha\beta}}{2} + \int_0^\infty d \omega \; \mbox{Re } \sigma_{\alpha\beta}^{(\rm regular)} (\omega) =\frac{\omega_{\rm p}^2}{8}\delta_{\alpha\beta} = \frac{\pi e^2 n}{2 m}\delta_{\alpha\beta} , \label{fsum} \] where $n = N/L^d$ is the electron density and $\omega_{\rm p}$ is the plasma frequency. 

Dissipation can be included phenomenologically in the Drude term by adopting a single-relaxation-time approximation, exactly as in the classical textbook case \cite{AM1,Allen06}, i.e. \[ \sigma_{\alpha\beta}^{(\rm Drude)}(\omega) = \frac{\tau}{\pi} \frac{D_{\alpha\beta}}{1 - i\omega\tau} , \label{tau} \] whose $\tau \rightarrow \infty$ limit coincides with the first term  in \equ{cond}.

In the special case of a band metal (i.e. a crystalline system of non interacting electrons) $\sigma_{\alpha\beta}^{(\rm regular)}(\omega)$ is a linear-response property which accounts for interband transitions, and is nonvanishing only at frequencies higher than a finite threshold; the threshold also survives after electron-electron interaction is turned on, owing to translational symmetry and the related selection rules.
In absence of translational symmetry the selection rule breaks down:  in disordered systems---and in disordered systems {\it only} \cite{Scalapino93}---$\sigma_{\alpha\beta}^{(\rm regular)}(0)$ may be nonzero (and the Drude weight may vanish).

Transverse conductivity is nonzero only when T-symmetry is absent. The Kubo formul\ae\  for the transverse (antiymmetric) conductivity are:
\bea \mbox{Re } \sigma_{\alpha\beta}^{(-)}(\omega) &=& \frac{2e^2}{\hbar  L^d} {\sum_{n\neq 0}} \frac{{\cal I}_{n,\alpha\beta}}{\omega_{0n}^2 - \omega^2} \label{s3}
\\ \mbox{Im } \sigma_{\alpha\beta}^{(-)}(\omega) &=& \frac{\pi e^2}{\hbar L^d} {\sum_{n
\neq 0}} \frac{{\cal I}_{n,\alpha\beta}}{\omega_{0n}} [  \delta(\omega - \omega_{0n}) -  \delta(\omega + \omega_{0n}) ] . \label{s4} \eea 

\section{Time-reversal even geometrical observables} \label{sec:even}

\subsection{Drude weight}

Electron transport in the diffusive regime is a balance between free acceleration and dissipation \cite{Allen06}; the Drude weight $D_{\alpha\beta}$ (also called adiabatic charge stiffness) is an intensive property of the pristine material, accounting for the former side of the phenomenon only.

In the case of a flat one-body potential (i.e. electron gas, either free or interacting)
the velocity operator $\hat{\v}$ is diagonal over the energy eigenstates: the matrix elements ${\cal R}_{n,\alpha\beta}$ in \equ{drude} vanish and $D_{\alpha\beta}$ assumes the same value as in classical physics \cite{Drude00,AM1}, i.e $D_{\alpha\beta} = \pi e^2 (n/m) \delta_{\alpha\beta}$.  Given \equ{fsum}, switching on the potential (one-body and two-body) has the effect of transferring some spectral weight from the Drude peak into the regular term. For free electrons the acceleration induced by a constant $\EEE$ field is $-e/m$, and the accelerating current is $-e$ times the mechanical acceleration.   $D_{\alpha\beta}$ measures then the free acceleration of the many-electron system induced by a field $\EEE$ constant in space, although in the adiabatic limit only (it is an $\omega=0$ linear response) \cite{rap157}; equivalently, it measures the (inverse) inertia of the electrons. \index{Drude} \index{conductivity}

The form of \equ{drude} does not explicitly show that $D_{\alpha\beta}$ is a ground-state property. In order to show that, I adopt the symbol ``$\doteq$'' with the meaning ``equal in the dc limit'', and I define $\sigma^{(\rm D)}_{\alpha\beta}(\omega) \doteq \partial j_\alpha(\omega)/\partial \EE_\beta(\omega)$. Conductivity requires the vector-potential gauge: we consider the response to a vector potential $A(\omega)$ in the dc limit: \[ \sigma^{(\rm D)}_{\alpha\beta}(\omega) \doteq  \frac{\partial j_\alpha}{\partial A_\beta} \frac{\partial A}{\partial \EE} . \] The $\kk$-dependent current was given above in \equ{current}; we notice that  \[ \frac{\partial j_\alpha}{\partial A_\beta} = \frac{e}{\hbar c} \frac{\partial j_\alpha}{\partial \kappa_\beta} = \left.  - \frac{e^2}{\hbar^2 c L^d} \frac{\partial^2 E_0(\kk)}{\partial \kappa_\alpha \partial \kappa_\beta} \right|_{\kk=0}\; \qquad \frac{dA}{d\EE} \doteq  -c \left[ \pi \delta(\omega) + \frac{i}{\omega} \right] , \] where the second expression comes from the causal inversion of $\EE(\omega) = i\omega A(\omega)/c$ \cite{Scalapino93}; we thus arrive at the famous Kohn's expression \cite{Kohn64}: \[ D_{\alpha\beta} = \frac{\pi e^2}{\hbar^2 L^d} \frac{\partial^2 E_0}{\partial \kappa_\alpha \partial \kappa_\beta}, \qquad  \sigma_{\alpha\beta}^{(\rm D)}(\omega) = D_{\alpha\beta} \left[ \delta(\omega) + \frac{i}{\pi \omega} \right]\label{kohnd} \] where we remind that it is crucial to set $\kk=0$ in the derivative before the large-$L$ limit is taken. From \equ{current} it is obvious that $D_{\alpha\beta}$ vanishes in insulators.

The expression in \equ{kohnd} is not yet geometrical; we arrive at an equivalent geometrical form starting from the identity $ \me{\Psi_{0\kk}}{\,( \hat{H}_{\kk} - E_{0\kk} )\,}{\Psi_{0\kk}} \equiv 0 \label{iden} $, taking two derivatives, and setting $\kk = 0$: \bea \frac{\partial^2 E_{0\kk}}{\partial \kappa_\alpha \partial \kappa_\beta} &=& \frac{N \hbar^2}{m} \delta_{\alpha\beta} - 2 \, \mbox{Re }
\, \me{\partial_{\kappa_\alpha} \Psi_{0\kk}}{\,( \hat{H}_{\kk} - E_{0\kk} )\,}{\partial_{\kappa_\beta} \Psi_{0\kk}} \label{Kohn3a} \\ D_{\alpha\beta} &=& \frac{\pi e^2 N}{m L^d} \delta_{\alpha\beta} - \frac{2 \pi e^2}{\hbar^2 L^d} \mbox{Re } \me{\partial_{\kappa_\alpha} \Psi_0}{\,( \hat{H} - E_0 )\,}{\partial_{\kappa_\beta} \Psi_0} , \label{Kohn3} \eea The two terms in \equ{Kohn3} have a very transparent meaning: the first one measures the free-electron acceleration; the geometrical term measures how much such acceleration is hindered by the one-body and two-body potentials. As observed above, the geometrical term is zero even for the {\it interacting} electron gas; whenever instead the one-body potential is not flat, then both one-body and two-body terms in $\hat{V}$  concur in hindering the free acceleration.

The geometrical term in \equ{Kohn3} can also be cast as a sum rule for longitudinal conductivity: from \equ{fsum} we have \[ \frac{\pi e^2}{\hbar^2 L^d} \mbox{Re } \me{\partial_{\kappa_\alpha} \Psi_0}{\,( \hat{H} - E_0 )\,}{\partial_{\kappa_\beta} \Psi_0} = \int_0^\infty d \omega \; \mbox{Re } \sigma_{\alpha\beta}^{(\rm regular)} (\omega) . \] On the experimental side, the partitioning of $\sigma_{\alpha\beta}^{(+)}(\omega)$ into a broadened Drude peak and a regular term $\sigma_{\alpha\beta}^{(\rm regular)}(\omega)$ is not so clearcut as one might wish \cite{Allen06}.

\subsection{Souza-Wilkens-Martin sum rule and the theory of the insulating state}

The insulating behavior of a generic material implies that  $D_{\alpha\beta}=0$ and that $\mbox{Re } \sigma_{\alpha\beta}^{(\rm regular)}(\omega)$ goes to zero for $\omega \rightarrow 0$ at zero temperature. For this reason Souza, Wilkens, and Martin (hereafter quoted as SWM) proposed to characterize the metallic/insulating behavior of a material by means of the integral \cite{Souza00} \[ I^{(\rm SWM)}_{\alpha\beta} = \int_0^\infty \frac{d \omega}{\omega} \mbox{Re } \sigma_{\alpha\beta}^{(+)} (\omega) \label{swm} , \] which diverges for all metals and converges for all insulators; in a gapped insulator the integrand is zero for $\omega < \epsilon_{\rm gap}/\hbar$. Owing to a fluctuation-dissipation theorem, the SWM integral is a geometrical property of the insulating ground state.

\subsubsection{Periodic boundary conditions} \label{sec:rs}

Dealing with dc conductivity obviously requires PBCs; whenever the Drude weight is nonzero, the integral in \equ{swm} diverges because of the $\delta(\omega)/\omega$ integrand. Therefore determining  whether $I^{(\rm SWM)}_{\alpha\beta}$ converges or diverges is completely equivalent to determining whether $D_{\alpha\beta}$ is zero or finite; it will be shown that the PBCs metric is related to $\sigma_{\alpha\beta}^{(\rm regular)}(\omega)$ only.

We insert a complete set of states into \equ{metric} at $\kk=0$ to obtain the intensive quantity: 
\[ {\mathfrak g}_{\alpha\beta} = \frac{1}{N} g_{\alpha\beta}(0) = \frac{1}{N} \mbox{Re }  \sum_{n \neq 0} \ev{\daa \Psi_0 | \Psi_n} \ev{\Psi_n | \dbb \Psi_0} . \label{sum2} \] We then evaluate the $\kk$-derivatives via perturbation theory in the parallel transport gauge: 
 \[ \ket{\partial_{\kappa_\alpha} \Psi_0} = - \sum_{n \neq 0} \ket{\Psi_n} \frac{\me{\Psi_n}{\hat v_\alpha }{\Psi_0}}{\omega_{0n}}, \qquad \qquad {\mathfrak g}_{\alpha\beta} = \frac{1}{N}\sum_{n \neq 0} \frac{\mbox{Re}\me{\Psi_n}{\hat v_\alpha }{\Psi_0}\me{\Psi_n}{\hat v_\beta}{\Psi_0}}{\omega_{0n}^2} \label{kdotp2} \] 

From the Kubo formula, \equ{s1}, we have \[ \int_0^\infty \frac{d \omega}{\omega} \mbox{Re } \sigma^{(\rm regular)}_{\alpha\beta} (\omega) =  \frac{\pi e^2}{\hbar L^d}  {\sum_{n\neq 0}} \frac{ {\cal R}_{n,\alpha\beta}}{\omega_{0n}^2} = \frac{\pi e^2 N}{\hbar L^d}  {\mathfrak g}_{\alpha\beta} , \label{metric2} \] where the $N \rightarrow \infty$ limit is understood. The intensive quantity ${\mathfrak g}_{\alpha\beta}$, having the dimensions of a squared length, in the case of a band insulator is related to the gauge-invariant quadratic spread $\Omega_{\rm I}$ of the Wannier functions \cite{Vanderbilt}: for an isotropic solid  \[ {\mathfrak g}_{xx} = \frac{\Omega_{\rm I}}{n_{\rm b}d} , \] where $n_{\rm b}$ is the number of occupied bands. It is seen from \equ{metric2} that ${\mathfrak g}_{\alpha\beta}$ does not discriminate between insulators and metals: it is finite in both cases. The story does not ends here, though.

In  1999 Resta and Sorella have defined a squared localization length $\lambda^2$ as a discriminant for the insulating state \cite{rap107}: as a function of $N$, $\lambda^2$ converges to a finite value in all insulators, and diverges in all metals. In the original paper the approach was demonstrated for the two-band Hubbard model of \equ{hub} and its quantum transition. Many years after the divergence/convergence of $\lambda^2$
has been successfully adopted for investigating the Mott transition in the paradigmatic case of a linear chain of hydrogen atoms \cite{Stella11}. In insulators $\lambda^2$ is a finite-$N$ approximant of ${\mathfrak g}_{xx}$, but when the same definition is applied to metals $\lambda^2$ has the virtue of diverging.
\index{Hubbard}
We assume an isotropic system and we consider once more $\kk_1 = (2\pi/L,0,0)$; since the metric is by definition the infinitesimal distance, \eqs{b1}{conne} yield to leading order \[ N {\mathfrak g}_{xx} \left( \frac{2\pi}{L} \right)^2 \simeq -  \mbox{ln } |\ev{\Psi_0|\Psi_{0\kk_1}}|^2 , \qquad  {\mathfrak g}_{xx} \simeq -\frac{L^2}{4\pi^2 N} \mbox{ln } |\ev{\Psi_0|\Psi_{0\kk_1}}|^2 . \label{rapxx} \]
If the system is insulating, we may replace $\ket{\Psi_{0\kk_1}}=\emi{\kk_1 \cdot \hat\r} \ket{{\Psi}_0}$ as we did in \equ{rap2} above:  \[ {\mathfrak g}_{xx} \simeq  -\frac{L^2}{4\pi^2 N} \mbox{ln } |\me{\Psi_0}{\ei{\frac{2\pi}{L} \sum_i x_i}}{\Psi_0}|^2 . \] The r.h.s. coincides indeed with $\lambda^2$, \equ{rs}, originally introduced in Ref \cite{rap107}. Given that ${\mathfrak g}_{xx}$ is intensive, the logarithm in \equ{rapxx} scales like $N^{1-2/d}$.

Next we address the metallic case. In a band metal $\ket{\Psi_0}$ is a Slater determinant of Bloch orbitals, and not all the $\k$ vectors in the Brillouin zone are occupied. A selection rule then guarantees that $\me{\Psi_0}{\ei{\frac{2\pi}{L} \sum_i x_i}}{\Psi_0}$ vanishes even at {\it finite} $N$ \cite{rap_a23,rap_a33}; therefore $\lambda^2$ is formally infinite. In disordered or correlated materials the selection rule breaks down, and $\lambda^2$ diverges in the large-$N$ limit only: this can be seen as follows. Whenever  the Drude weight is nonzero, then \equ{kohnd} guarantees that $\ket{\Psi_{0\kk_1}}$ is an eigenstate of ${\hat H}_{\kk_1}$ orthogonal to $\emi{\kk_1 \cdot \hat\r} \ket{{\Psi}_0}$; to lowest order in $\kk_1$ we have: \[ 0 = \me{\Psi_{0}}{\ei{\kk_1 \cdot \hat\r}}{\Psi_{0\kk_1}} \simeq  \me{\Psi_{0}}{\ei{\kk_1 \cdot \hat\r}}{\Psi_0} , \label{ortho} \] which proves the divergence of $\lambda^2$. In the large-$L$ limit the matrix element's modulus |$\me{\Psi_{0}}{\ei{\kk_1 \cdot \hat\r}}{\Psi_0}|$ approaches one from below in insulators, while it approaches zero in metals.

\subsubsection{Open boundary conditions}

The SWM integral is more useful in practical computations within OBCs. A bounded sample does not support a dc current, and $D_{\alpha\beta} = 0$ at any finite size: this is consistent with the fact that \equ{current} vanishes within OBCs. An oscillating field $\EEE(\omega)$ in a large sample linearly induces a macroscopic polarization ${\bf P}(\omega)$; since ${\bf j}(t) = d {\bf P}(t)/dt$, we define a ``fake'' conductivity by means of the relationship \[ \tilde\sigma_{\alpha\beta}(\omega) = -i\omega \frac{\partial P_\alpha(\omega)}{\partial \EE_\beta(\omega)} . \]

The Kubo formul\ae\ for this OBCs response function are: \[ \mbox{Re } \tilde\sigma_{\alpha\beta}(\omega) = \frac{\pi e^2}{\hbar L^d}  {\sum_{n\neq 0}} \frac{ {\cal R}_{n,\alpha\beta}}{\omega_{0n}} [\, \delta(\omega - \omega_{0n}) + \delta(\omega + \omega_{0n}) \,] . \] Despite the formal similarity with \equ{s1}, $\tilde\sigma_{\alpha\beta}(\omega)$ is {\it very different}---at finite size---from $\sigma_{\alpha\beta}^{(\rm regular)}(\omega)$: different eigenvalues, different matrix elements and selection rules; also, $\tilde\sigma_{\alpha\beta}(\omega)$ saturates the $f$-sum rule, while $\sigma_{\alpha\beta}^{(\rm regular)}(\omega)$ by itself does not (in metals). Then it is easy to show that the SWM integral is related to the OBCs metric in the same way as in \equ{metric2}:  \[ \int_0^\infty \frac{d \omega}{\omega} \mbox{Re } \tilde\sigma_{\alpha\beta} (\omega) =  \frac{\pi e^2}{\hbar L^d}  {\sum_{n\neq 0}} \frac{ {\cal R}_{n,\alpha\beta}}{\omega_{0n}^2} = \frac{\pi e^2 N}{\hbar L^d}  \tilde{\mathfrak g}_{\alpha\beta} , \label{metric3} \] where again the $N \rightarrow \infty$ limit is understood. The OBCs metric per electron $\tilde{\mathfrak g}_{\alpha\beta}$ coincides with the PBCs one in insulators, but has the virtue of diverging in metals \cite{rap132}. What actually happens is that the low-frequency spectral weight in the OBCs $\tilde\sigma_{\alpha\beta}(\omega)$ is reminiscent of---and accounts for---the corresponding Drude peak within PBCs, thus leading to a diverging $I^{(\rm SWM)}_{\alpha\beta}$.

Within OBCs one has $\ket{\partial_{\kk} \Psi_0} = -i \hat{\r} \ket{\Psi_0}$, hence \equ{metric} yields \[ \tilde{\mathfrak g}_{\alpha\beta} = \frac{1}{N} ( \, \me{\Psi_0}{\hat{r}_\alpha \hat{r}_\beta}{\Psi_0} - \me{\Psi_0}{\hat{r}_\alpha}{\Psi_0} \me{\Psi_0}{\hat{r}_\beta}{\Psi_0} \,), \] and the ``Re'' is not needed. This is clearly a second cumulant moment of the dipole (per electron): the symbol $\ev{r_\alpha r_\beta}_{\rm c}$ has been equivalently used in some previous literature. 
Alternatively, $\tilde{\mathfrak g}_{\alpha\beta}$ measures the quadratic quantum fluctuations of the polarization in the ground state \cite{Souza00}. An equivalent expression for $\tilde{\mathfrak g}_{\alpha\beta}$ is in terms of the one-body density $n(\r)$ and the two-body density $n^{(2)}(\r,\r')$ \cite{rap132}: \bea \tilde{\mathfrak g}_{\alpha\beta} &=& \frac{1}{2N} \int d {\bf r}\,  d {\bf r}'\;({\bf
r} - {\bf r}')_\alpha ({\bf r} - {\bf r}')_\beta  [\,  n({\bf
r}) n({\bf r}') - n^{(2)}({\bf r},{\bf r}') \, ] .  \label{prov4} \nn &=& - \frac{1}{2N} \int d {\bf r}\,  d {\bf r}'\;({\bf
r} - {\bf r}')_\alpha ({\bf r} - {\bf r}')_\beta \; n({\bf r}) \,
 n_{\rm xc}({\bf r},{\bf r}') , \label{xc} \eea where $ n_{\rm xc}({\bf r},{\bf r}')$ is by definition the exchange-correlation hole density.
Therefore $\tilde{\mathfrak g}_{\alpha\beta}$ is a
second moment of the exchange-correlation hole, averaged over the sample.

Finite-size model-Hamiltonian OBCs calculations have provided---by means of $\tilde{\mathfrak g}_{\alpha\beta}$---insight into the Anderson insulating state in 1$d$ \cite{rap143,rap156}, and into the Anderson metal-insulator transition in 3$d$ \cite{rap152}.
  
\section{Time-reversal odd geometrical observables} \label{sec:odd}

\subsection{Anomalous Hall conductivity and Chern invariant} \index{Hall effect}

Edwin Hall discovered the eponymous effect in 1879; two years later he discovered the anomalous Hall effect in ferromagnetic metals. The latter is, by definition, the Hall effect in absence of a macroscopic ${\bf B}$ field. Nonvanishing transverse conductivity requires breaking of T-symmetry: in the normal Hall effect the symmetry is broken by the applied ${\bf B}$ field; in the anomalous one it is spontaneously broken, for instance by the development of ferromagnetic order. The theory of anomalous Hall conductivity  in metals has been controversial for many years; since the early 2000s it became clear that, besides extrinsic effects, there is also an intrinsic contribution, which can be expressed as a geometrical property of the electronic ground state in the pristine crystal. 
\index{curvature} \index{topological invariant}
Without extrinsic mechanisms the longitudinal dc conductivity would be infinite; such mechanisms are necessary to warrant Ohm's law, and are accounted for by relaxation time(s) $\tau$; in absence of T-symmetry, extrinsic mechanisms affect the anomalous Hall conductivity (AHC) as well. Two distinct mechanisms have been identified: they go under the name of ``side jump'' and ``skew scattering'' \cite{Nagaosa10}.  The side-jump term is nondissipative (independent of $\tau$). Since a crystal with impurities actually is a (very) dilute alloy, we argued that the sum of the intrinsic and side-jump terms can be regarded as the intrinsic term of the alloy \cite{rap149}. As a matter of principle, such ``intrinsic'' AHE of the dirty sample can be addressed either in reciprocal space \cite{rap149,rap135}, or even in real space \cite{rap153}. Finally the skew-scattering term is dissipative, proportional to $\tau$ in the single-relaxation-time approximation. Here we deal with the intrinsic geometrical term only.

As pointed out by Haldane in a milestone paper appeared in 1988 \cite{Haldane88}, AHC is also allowed in insulators, and is topological in 2$d$: therein extrinsic effects are ruled out. In fact in insulators the dc longitudinal conductivity is zero, and---as a basic tenet of topology---any impurity has no effect on the Hall conductivity insofar as the system remains insulating. The effect goes under the name of quantum anomalous Hall effect (QAHE); the synthesis of a 2$d$ material where the QAHE occurs was only achieved since 2013 onwards \cite{Chang13,Chang15}.

 The Kubo formula of \equ{s3} immediately gives the intrinsic AHC term as:
\[ \mbox{Re } \sigma_{\alpha\beta}^{(-)}(0) = \frac{2e^2}{\hbar  L^d} {\sum_{n\neq 0}}' \frac{\mbox{Im }\langle \Psi_0 | \hat v_\alpha | \Psi_n \rangle \langle \Psi_n | \hat v_\beta | \Psi_0 \rangle}{\omega_{0n}^2} . \] In a similar way as for \eqs{kdotp2}{metric2}, we easily get \[ \mbox{Re } \sigma_{\alpha\beta}^{(-)}(0) = - \frac{e^2}{\hbar  L^d} {\sf \Omega}_{\alpha\beta}(0) , \] where ${\sf \Omega}_{\alpha\beta}(0)$ is the many-body Berry curvature, \equ{curva}. The expression holds for metals and insulators, in either 2$d$ or 3$d$; the large-system limit is understood. In the band-structure case ${\sf \Omega}_{\alpha\beta}(0)/L^d$  is simply related to the Fermi-volume integral of the one-body Berry curvature \cite{Vanderbilt,rap135}. 

 Next we consider the 2$d$ case: in any smooth gauge the curvature per unit area can be written as 
  \[ \frac{1}{L^2} {\sf \Omega}_{xy}(0)   =   \frac{1}{L^2}  \left( \frac{L}{2\pi} \right)^2 \int_0^{\frac{2\pi}{L}} \!\! d\kappa_x \int_0^{\frac{2\pi}{L}} \!\! d\kappa_y \; {\sf \Omega}_{xy}(\kk)  , \label{ch1}\] where the integral and the prefactor are both dimensionless. Even this formula holds for both insulators and metals, 
 but we remind that $\ket{\Psi_{0\kk}}$ obtains by following  $\ket{\Psi_{0}}$ as the flux $\kk$ is adiabatically turned on; the behavior of the integrand in \equ{ch1} is then {\it qualitatively} different in the insulating vs. metallic case. 
 
We deal with the insulating case only from now on. Since the Drude weight is zero, the energy is $\kk$ independent; furthermore the integral in \equ{ch1} is actually equivalent to the integral over a torus and is therefore quantized. In order to show this, we observe that
whenever the components of $\kk - \kk'$ are integer multiples of $2\pi/L$, then the state $\ei{(\kk-\kk')\cdot \hat{\r}} \ket{\Psi_{0\kk}}$ is eigenstate of $\hat{H}_{\kk'}$ with the same eigenvalue: \[ \ket{\Psi_{0\kk'}} = \ei{(\kk-\kk')\cdot \hat{\r}} \ket{\Psi_{0\kk}} . \label{gauge} \] Since ${\sf \Omega}_{xy}(\kk)$ is gauge-invariant, an arbitrary phase factor may relate the two members of \equ{gauge}. It is worth stressing that in the topological case a globally smooth gauge does not exist; in other words we can enforce \equ{gauge} as it stands (with no extra phase factor) only locally, not globally.\footnote{The gauge choice of \equ{gauge} is the many-body analogue of the periodic gauge in band-structure theory. Therein it is well known that, in the topological case, it is impossible to adopt a gauge which is periodic and smooth on the whole Brillouin zone: an ``obstruction'' is necessarily present. See Ch. 3 in Ref. \cite{Vanderbilt}.} \index{Hall effect}
 
The integral in \equ{ch1} is quantized (even at finite $L$) and is proportional to the many-body Chern number, as defined by Niu, Thouless and Wu (NTW) in a famous paper \cite{Niu85}: \[ C_1 = \frac{1}{2\pi} \int_0^{\frac{2\pi}{L}} \!\! d\kappa_x \int_0^{\frac{2\pi}{L}} \!\! d\kappa_x  \;  {\sf \Omega}_{xy}(\kk)  , \qquad \mbox{Re } \sigma_{xy}^{(-)}(0) = - \frac{e^2}{h} C_1 .\label{ch2}  \] In the present formulation we have assumed PBCs at any $\kk$, and---following Kohn \cite{Kohn64}---we have ``twisted'' the Hamiltonian. The reverse is done by NTW: the Hamiltonian is kept fixed, and the boundary conditions are ``twisted''. It is easy to show that the two approaches are equivalent: within both of them the two components of $\kk$ become effectively angles, the integration is over a torus, and the integral is a topological invariant.
 
 The AHC is therefore quantized in any 2$d$ T-breaking insulator, thus yielding the QAHE. Originally, NTW were not addressing the QAHE; the phenomenon addressed was instead the fractional quantum Hall effect, where the electronic ground state is notoriously highly correlated \cite{Laughlin83f}. While the topological invariant is by definition integer, the fractional conductance owes---according to NTW---to the degeneracy of the ground state in the large-$L$ limit. Also, in presence of a macroscopic ${\bf B}$ field, gauge-covariant boundary conditions and magnetic translations must be adopted.

\subsection{Magnetic circular dichroism sum rule}

Since a very popular (and misleading) paper appeared in 1992 \cite{Thole92}, magnetic circular dichroism (MCD) has been widely regarded among synchrotron experimentalists as an approximate probe of orbital magnetization $\M$ in bulk solids. It became clear over the years that this is an unjustified assumption, thanks particularly to Refs. \cite{Kunes00}, \cite{Souza08}, and \cite{rap159}.
\index{dichroic spectroscopy}
The differential absorption of right and left circularly polarized light by magnetic materials is known as magnetic circular dichroism; the object of interest is 
the frequency integral of the imaginary part of the antisymmetric term in the conductivity tensor \[ I_{\alpha\beta}^{(\rm MCD)} = \mbox{Im } \int_0^\infty d\omega \; \sigma^{(-)}_{\alpha\beta}(\omega) ; \label{sum} \] a kind of fluctuation-dissipation theorem relates $I_{\alpha\beta}^{(\rm MCD)}$ to a ground-state property. The Kubo formula, \equ{s4}, immediately yields 
\[ I_{\alpha\beta}^{(\rm MCD)} =  \frac{\pi e^2}{\hbar L^d} {\sum_{n
\neq 0}} \frac{{\cal I}_{n,\alpha\beta}}{\omega_{0n}}  = \frac{\pi e^2}{\hbar L^3} \mbox{Im}\sum_{n\neq 0} \frac{ \me{\Psi_0}{{\hat v}_\alpha}{\Psi_n}\me{\Psi_n}{{\hat v}_\beta}{\Psi_0}}{\omega_{0n}} ; \label{corr}\] this expression holds both within OBCs and PBCs, although with different eigenvalues, different matrix elements, and different selection rules (at any finite size). In both cases, $I_{\alpha\beta}^{(\rm MCD)}$ can be cast as a geometric property of the electronic ground state, via the substitution  \[ \ket{\partial_{\kk} \Psi_0} = - \sum_{n\neq 0} \ket{\Psi_n} \frac{\me{\Psi_n}{\hat{\bf v}}{\Psi_0}}{\omega_{n0}} \] \[ (\hat{H} - E_0) \ket{\partial_{\kk} \Psi_0} = - \sum_{n\neq 0} \ket{\Psi_n} \me{\Psi_n}{\hat{\bf v}}{\Psi_0} . \] By comparing the last expression to \equ{corr} the geometrical formula is
\[ I_{\alpha\beta}^{(\rm MCD)} = \frac{\pi e^2}{\hbar^2 L^3} \mbox{Im } \me{\partial_{\kappa_\alpha} \Psi_0 }{(\hat{H} - E_0)}{\partial_{\kappa_\beta} \Psi_0} . \label{cmd} \]

The PBCs many-body expression for $I_{\alpha\beta}^{(\rm MCD)}$, \equ{cmd}, unfortunately cannot be compared with a corresponding formula for $\M$. To this day such a formula does not exist: the orbital magnetization of a correlated many-body wavefunction within PBCs is currently an open (and challenging) problem. A thorough comparison has been done at the band-structure level only, where both $I_{\alpha\beta}^{(\rm MCD)}$ and $\M$ have a known expression, as a Fermi-volume integral of a geometrical integrand \cite{Souza08,rap159}.

A direct comparison between $I_{\alpha\beta}^{(\rm MCD)}$ and $\M$ was instead provided within OBCs as early as 2000 by Kunes and Oppeneer \cite{Kunes00}; we are going to retrieve their outstanding result within the present formalism. As already observed, within OBCs one has $\ket{\partial_{\kk} \Psi_0} = -i \hat{\r} \ket{\Psi_0}$, ergo \bea I_{\alpha\beta}^{(\rm MCD)} &=&  \frac{\pi e^2}{\hbar^2 L^3} \mbox{Im } \me{\Psi_0 }{\hat{r}_\alpha(\hat{H} - E_0)\hat{r}_\beta}{\Psi_0} =  - \frac{i \pi e^2}{2\hbar^2 L^3} \mbox{Im } \me{\Psi_0 }{\hat{r}_\alpha [\hat{H},\hat{r}_\beta]}{\Psi_0} \nn &=& - \frac{\pi e^2}{2\hbar L^3}  \me{\Psi_0 }{(\hat{r}_\alpha\hat{v}_\beta -\hat{r}_\beta\hat{v}_\alpha)}{\Psi_0} . \label{kunes} \eea 

The ground-state expectation value in \equ{kunes} was originally dubbed ``center of mass angular momentum''. By expanding the many-body operators $\hat{\r}$ and $\hat{{\bf v}}$, the matrix element is the ground-state expectation value of $\sum_{ii'} \r_i \times {\bf v}_{i'}$, while the orbital moment of a bounded sample is proportional to the expectation value of $\sum_{i} \r_i \times {\bf v}_{i}$. The two coincide only in the single-electron case; this is consistent with the band-structure findings. Indeed it has been proved that $I_{\alpha\beta}^{(\rm MCD)}$ and $\M$ coincide {\it only} for an isolated flat band: a disconnected electron distribution with one electron per cell (and per spin channel) \cite{rap159}.

The MCD sum rule $I_{\alpha\beta}^{(\rm MCD)}$ is an outstanding ground-state observable per se, because of reasons not to be explained here, and which are at the root if its enormous experimental success. Notwithstanding, there is no compelling reason for identifying it, even approximately, with some form of orbital magnetization. The two observables $I_{\alpha\beta}^{(\rm MCD)}$ and $\M$ provide a quantitatively different measure of spontaneous T-breaking in the orbital degrees of freedom of a given material.

\section{Conclusions} \label{sec:conclu}

The known geometrical observables come in two very different classes: those in class (i) only make sense for insulators, and are defined modulo $2\pi$ (in dimensionless units), while those in class (ii) are defined for both insulators and metals, and are single-valued. Such outstanding difference owes---at the very fundamental level---to the fact that the observables in class (i) are expressed by means of gauge-dependent $(2n-1)$-forms, while those in class (ii) are expressed in terms of gauge-invariant $2n$-forms.

I have thoroughly discussed here the only observable in class (i) whose many-body formulation is known: macroscopic polarization \cite{rap100}; it is rooted into the Berry connection, a gauge-dependent 1-form, called Chern-Simons 1-form in mathematical speak. 
The many-body connection, \equ{conne}, may yield a physical observable only after the gauge is fixed: in the present case, I adopted the many-body analogue of the periodic gauge in band-structure theory.
I have also discussed the multivalued nature of bulk polarization, whose features in either 1$d$ or 3$d$ are somewhat different. 

Another class-(i) geometrical observable is known in band-structure theory, where it is expressed as the Brillouin-zone integral of a Chern-Simons 3-form: this is the so called ``axion'' term in magnetoelectric response \cite{Vanderbilt}. The corresponding many-body expression is not known; it is even possible that it could not exist as a matter of principle \cite{rapix2}.
In presence of some protecting symmetry a class-(i) observable may only assume the values zero or $\pi$ (mod $2\pi$): the observable becomes then a topological  ${\mathbb Z}_2$ index: a ${\mathbb Z}_2$-odd crystalline insulator cannot be ``continuously deformed'' into a ${\mathbb Z}_2$-even without passing through a metallic state and without breaking the protecting symmetry. 

Four geometrical observables of class (ii), having a known many-body expression, have been discussed in the present Review. All are single valued, and all are rooted in gauge-invariant 2-forms; the following table summarizes them:

\begin{center} \begin{tabular}{|c|c|} \hline 
{\bf Time-reversal odd} & {\bf Time-reversal even} \\ 
\hline  \hline
Anomalous Hall conductivity & Souza-Wilkens-Martin sum rule  \\ \hline
Magnetic circular dichroism sum rule & Drude weight \\ \hline \end{tabular} \end{center}

The four observables are expressed by means of the $\kk=0$ values of the geometrical  2-forms $\FF$ and $\GG$, defined as follows: \bea \FF &=&  [ \, \ev{\partial_{\kappa_\alpha} \Psi_{0\kk} | \partial_{\kappa_\beta} \Psi_{0\kk}} - \ev{\partial_{\kappa_\alpha} \Psi_{0\kk} | \Psi_{0\kk} } \ev{\Psi_{0\kk} | \partial_{\kappa_\beta} \Psi_{0\kk}}\,] \,d\kappa_\alpha d\kappa_\beta , \\ \GG &=&  \me{\partial_{\kappa_\alpha} \Psi_{0\kk}}{\,( \hat{H}_{\kk} - E_{0\kk} )\,}{\partial_{\kappa_\beta} \Psi_{0\kk}} \, d\kappa_\alpha d\kappa_\beta .
\eea Both forms are extensive; the real symmetric part of $\FF$ coincides with the quantum metric, \equ{metric}, while its imaginary part (times $-2$) coincides with the Berry curvature, \equ{curva}.

The T-odd observables in the table obtain from the antisymmetric imaginary part of $\FF$ (AHC) and $\GG$ (MCD sum rule); similarly, the T-even observables obtain from the real symmetric part of $\FF$ (SWM sum rule) and $\GG$ (Drude weight). All of these observables have an elegant independent-electron crystalline counterpart: within band-structure theory they are expressed as Fermi volume integrals (Brillouin-zone integrals in insulators) of gauge-invariant geometrical 2-forms in Bloch space \cite{rapix}. There is one very important T-odd geometrical observable missing from the above table: orbital magnetization. Its expression within band-structure theory is known since 2006, for both insulators and metals \cite{Vanderbilt,rap130}. Very baffingly, a corresponding expression in terms of the many-body ground state does not exist to this day.

The two T-even observables have an important meaning in the theory of the insulating state \cite{rap_a33}. The Drude weight is zero in all insulators, and nonzero in all metals; to remain on the safe side, the statement applies to systems without disorder \cite{Scalapino93}. Instead the geometrical term in the SWM sum rule within PBCs does not discriminate between insulators and metals; nonetheless I have shown that a discretized formulation of the same observable---proposed by Resta and Sorella back in 1999 \cite{rap107}---does discriminate. Furthermore it is expedient to alternatively cast the SWM sum rule in the OBCs Hilbert space: even in this case the geometrical observable acquires the virtue of discriminating between insulators and metals \cite{rap_a33,rap152}.

Among the five observables dealt with in this Review, only two may become topological. I have shown that the polarization of a 1$d$ (or quasi-1$d$) inversion-symmetric insulator is a topological ${\mathbb Z}_2$ invariant (in electron-charge units): Fig. \ref{fig:quantum2} perspicuously shows that polyacetylene is a ${\mathbb Z}_2$-even topological case.
In modern jargon, the ${\mathbb Z}_2$ invariant is ``protected'' by inversion symmetry. Notably, the---closely related---topological nature of the soliton charge in polyacetylene was discovered long ago \cite{Su79}. The second geometrical observable which may become topological is the AHC: this occurs in 2$d$ insulators whenever T-symmetry is absent. Therein the AHC in natural conductance units\footnote{In 2$d$ conductivity and conductance have the same dimensions; their natural units are klitzing$^{-1}$. The klitzing (natural resistance unit) is defined as $h/e^2$.} is a ${\mathbb Z}$ invariant (Chern number): the effect is known as QAHE (quantum anomalous Hall effect) \cite{Chang13,Chang15}. The same ${\mathbb Z}$ invariant plays the key role in the theory of the fractional quantum Hall effect \cite{Niu85}.

\section*{Acknowledgments} During several stays at the Donostia International Physics Center in San Sebastian (Spain) over the years, I have discussed thoroughly the topics in the present Review with Ivo Souza; his invaluable contribution is gratefully acknowledged. The hospitality by the Center is acknowledged as well. Work supported by the ONR Grant No. No. N00014-17-1-2803.

\section*{Appendix}
\appendix

As in the main text we address a simple cubic lattice of constant $a$, with $L=Ma$; here we consider the electronic term only. We define $[\r] = (\r_1,\r_2\dots\r_N)$, and we indicate with the simple integral symbol $\int$ a multidimensional integral over the segment $(0,L)$ in each variable. Then the integral over the hypercube is
\bea \me{\Psi_0}{\ei{\frac{2\pi}{L}\, \sum_i x_i }}{\Psi_0} &=& \int \prod_{i=1}^N d \r_i  \; \ei{\frac{2\pi}{L}\, \sum_i x_i } |\ev{[\r]|\Psi_0}|^2  \\ &=& 
\int dy_1 dz_1 \int dx_1 \prod_{i=2}^N d \r_i  \; \ei{\frac{2\pi}{L}\, \sum_i x_i } |\ev{[\r]|\Psi_0}|^2 . \label{hyper} \eea Under the crystalline hypothesis, the inner integral is a lattice-periodical function of $(y_1,z_1)$, hence \[ \me{\Psi_0}{\ei{\frac{2\pi}{L}\, \sum_i x_i }}{\Psi_0} = M^2 \int_{\rm cell} \!\!\! dy_1 dz_1 \int dx_1 \prod_{i=2}^N d \r_i  \; \ei{\frac{2\pi}{L}\, \sum_i x_i } |\ev{[\r]|\Psi_0}|^2 . \] As defined in the main text, the reduced Berry phase is then \[ \tilde\gamma_x^{(\rm el)} = \mbox{Im ln } \int_{\rm cell} \!\!\! dy_1 dz_1 \int dx_1 \prod_{i=2}^N d \r_i  \; \ei{\frac{2\pi}{L}\, \sum_i x_i } |\ev{[\r]|\Psi_0}|^2 . \] This holds for a correlated wavefunction in a perfect lattice; in case of chemical disorder one instead averages over the disorder by evaluating \equ{hyper} on the large supercell and then dividing it by $M^2$ before taking the ``Im ln''. A similar reasoning applies to the nuclear term as well: hence \equ{reduced} in the main text.

\index{band structure} \index{Berry phase}
 At the independent-electron level $\ket{\Psi_0}$ is the Slater determinant of $N$ Bloch orbitals. We get rid of trivial factors of 2 by addressing spinless electrons; furthermore we consider the contribution to $P_x^{(\rm el)}$ of a single occupied band. The $\k_m$ Bloch vectors are: \[ m \equiv (m_1,m_2,m_3), \qquad \k_m = \frac{2\pi}{L}(m_1,m_2,m_3), \quad m_s = 0,1,\dots,M-1. \] The Bloch orbitals $\ket{\psi_{\k_m}} = \ei{\k_m \cdot \r} \ket{u_{\k_m}}$ are normalized over the crystal cell of volume $a^3$. It is expedient to define the auxiliary Bloch orbitals $\ket{\tilde\psi_{\k_m}} = \ei{\frac{2\pi}{L}\, x } \ket{\psi_{\k_m}}$, and $\ket{\tilde\Psi_0}$ as their Slater determinant; we also define ${\bf q}= (2\pi/L,0,0)$. Then \[ \me{\Psi_0}{\ei{\frac{2\pi}{L}\, \sum_i x_i }}{\Psi_0} =\me{\Psi_0}{\ei{ \sum_i{\bf q} \cdot  \r_i }}{\Psi_0}   = \ev{\Psi_0|\tilde\Psi_0} = \frac{1}{M^{3N}}\mbox{det } {\cal S} , \] where ${\cal S}$ is the $N \times N$ overlap matrix, in a different normalization: \bea {\cal S}_{mm'} &=& M^3 \ev{\psi_{\k_m}|\tilde \psi_{\k_{m'}}} = M^3\me{u_{\k_m}}{\ei{({\bf q} + \k_{m'} - \k_m) \cdot \r}}{u_{\k_{m'}}} 
 \nn &=&  M^3 \ev{u_{\k_m} | u_{\k_{m'}}} \, \delta_{{\bf q} + \k_{m'} - \k_m} = M^3 \ev{u_{\k_m}|u_{\k_{m}-{\bf q}}}  \delta_{mm'}.
\eea The normalization factors cancel: we have in fact \[ \me{\Psi_0}{\ei{\frac{2\pi}{L}\, \sum_i x_i }}{\Psi_0}  = \frac{1}{M^{3N}}\mbox{det } {\cal S} = \prod_{m_1,m_2,m_3 =0}^{M-1} \ev{u_{\k_m}|u_{\k_{m}-{\bf q}}} , \] \[ \tilde\gamma_x^{(\rm el)} = \frac{1}{M^2} \sum_{m_2,m_3 =0}^{M-1} \mbox{Im ln } \prod_{m_1 =0}^{M-1} \ev{u_{\k_m}|u_{\k_{m}-{\bf q}}} ; \] the multi-band case is dealt with in detail in Ref. \cite{rap_a30}.

 

\clearchapter

\end{document}